\documentclass{article}[12pt]
\usepackage{etoolbox}
\usepackage[utf8]{inputenc}
\usepackage[top = 2 cm, bottom = 2 cm, left = 2.5 cm, right = 1.5 cm]{geometry}
\usepackage[]{amsmath}
\usepackage[]{amssymb}
\usepackage[pdftex]{graphicx}
\graphicspath{ {images/} }
\usepackage{float}
\restylefloat{figure}
\usepackage{expl3}
\usepackage{subfig}
\usepackage{indentfirst}
\usepackage{amsthm}
\usepackage{amsfonts}
\usepackage{ mathrsfs }
\usepackage{mathtools}
\usepackage{cmap}
\usepackage{exercise}
\usepackage{tikz}
\usetikzlibrary{tikzmark,calc}
\usetikzlibrary{decorations.pathmorphing}
\usepackage{breqn}
\usepackage{cancel}
\usepackage{comment}
\usepackage{amsfonts,amssymb}
\usepackage{xcolor}
\usepackage{tcolorbox}
\usepackage{url}
\definecolor{mintcream}{rgb}{0.96, 1.0, 0.98}
\definecolor{champagne}{rgb}{0.97, 0.91, 0.81}
\definecolor{bubblegum}{rgb}{0.99, 0.76, 0.8}
\usepackage[offset=1.2em]{simpler-wick}
\usepackage[compat=1.1.0]{tikz-feynman}
\usepackage{tikz-3dplot}

\usepackage[pdftex,unicode,
            colorlinks=true,
          linkcolor = blue,
          citecolor = cyan,
          urlcolor = bubblegum]{hyperref}
\usepackage[colorinlistoftodos]{todonotes}

\newcommand{\Tr}{\operatorname{Tr}}

\newcommand{\htau}{\hat{\tau}}
\newcommand{\tauh}{\hat{\tau}}
\newcommand{\taub}{\bar{\tau}}

\def\be{\begin{eqnarray}}
\def\ee{\end{eqnarray}}
\def\nn{\nonumber}

\def\p{\partial}

\def\Tr{{\rm Tr}\,}

\usepackage{marginnote}

\definecolor{nicklethistwink}{rgb}{0.69, 0.19, 0.38}
\definecolor{ticklemepink}{rgb}{0.99, 0.54, 0.67}
\definecolor{mistyrose}{rgb}{1.0, 0.89, 0.88}
\definecolor{lolololol}{rgb}{0.25, 0.5, 0.75}
\definecolor{upsdellred}{rgb}{0.68, 0.09, 0.13}
\definecolor{ghostwhite}{rgb}{0.97, 0.97, 1.0}
\definecolor{ivory}{rgb}{1.0, 1.0, 0.94}
\definecolor{bubblegum}{rgb}{0.99, 0.76, 0.8}
\definecolor{honeydew}{rgb}{0.94, 1.0, 0.94}
\definecolor{magnolia}{rgb}{0.97, 0.96, 1.0}
\definecolor{isabelline}{rgb}{0.96, 0.94, 0.93}
\definecolor{bubbles}{rgb}{0.91, 1.0, 1.0}
\definecolor{floralwhite}{rgb}{1.0, 0.98, 0.94}

\begin{document}

\begin{center}
\begin{small}
\hfill FIAN/TD-18/22\\
\hfill ITEP/TH-20/22\\
\hfill IITP/TH-19/22\\
\hfill MIPT/TH-17/22\\

\end{small}
\end{center}

\vspace{.5cm}

\begin{center}
\begin{Large}\fontfamily{cmss}
\fontsize{15pt}{27pt}
\selectfont
	\textbf{AGT correspondence, (q-)Painlev\`e equations and matrix models}
	\end{Large}
	
\bigskip \bigskip

\begin{large}
A. Mironov$^{a,b,c,}$\footnote{mironov@lpi.ru; mironov@itep.ru},
V. Mishnyakov$^{d,a,c,e,}$\footnote{mishnyakovvv@gmial.com},
A. Morozov$^{d,b,c,}$\footnote{morozov@itep.ru},
Z. Zakirova$^{c,f,}$\footnote{zolya\_zakirova@mail.ru}
\end{large}

\bigskip

\begin{small}
$^a$ {\it Lebedev Physics Institute, Moscow, Russia}\\
$^b$ {\it Institute for Information Transmission Problems, Moscow, Russia}\\
$^c$ {\it NRC ``Kurchatov Institute" - ITEP, Moscow, Russia}\\
$^d$ {\it MIPT, Dolgoprudny, Russia}\\
$^e$ {\it Institute for Theoretical and Mathematical Physics, Lomonosov Moscow State University, Moscow, Russia}\\
$^f$ {\it Kazan State Power Engineering University, Kazan, Russia}
\end{small}
 \end{center}

\bigskip

\begin{abstract}
Painlev\`e equation for conformal blocks is a combined corollary of integrability and Ward identities,
which can be explicitly revealed in the matrix model realization of AGT relations.
We demonstrate this in some detail, both for $q$-Painlev\`e equations for the $q$-Virasoro conformal block, or AGT dual gauge theory in $5d$, and for ordinary Painlev\`e equations, or AGT dual gauge theory in $4d$.
Especially interesting is the continuous limit from $5d$ to $4d$ and its description at the level of
equations for eight $\tau$-functions.
Half of these equations are governed by integrability and another half by Ward identities.
\end{abstract}

\bigskip

\section{Introduction}

AGT relations \cite{AGT}
identify LMNS integrals \cite{LMNS}, or Nekrasov functions \cite{Nek} and $2d$ conformal blocks \cite{BPZ}.
They are best understood \cite{MMSh} as a Hubbard-Stratanovich duality
in the Dotsenko-Fateev \cite{DF} matrix model \cite{MMSh1,AGTmamo}
(which belongs to the class of Penner type models with logarithmic potential).
An amusing corollary \cite{Gamayun:2013auu} (see also later development in \cite{GIL})
is that a peculiar linear combination of conformal blocks
satisfies the Panl\`eve VI equation \cite{Plist},
which {\it a priori} seems to have nothing to do with any of the ingredients of
Nekrasov/AGT/matrix model theory.
Moreover, this Painlev\`e equation turns out to result from the two complementary
features of matrix models \cite{UFN3}: integrability and Ward/Virasoro identities,
which are alternatively combined into a {\it superintegrability} property \cite{MMsi}
of these models, which is now understood \cite{MMNek}
to be the true origin of Nekrasov calculus.
This can imply a more direct connection between superintegrability and Painlev\`e,
which, however, needs to be investigated better
and stays beyond the scope of the present paper.

The Dotsenko-Fateev (DF) matrix model is not Gaussian,
therefore it has numerous Dijkgraaf-Vafa phases,
differing by the choice of integration contours.
The problem is that for a particular choice of the phase
(unless all the contours are just the same), the
partition function does not possess
a determinant representation, and is not a KP $\tau$-function.
Instead, $\tau$-functions arise
as a simple linear combinations of DF integrals \cite{MMZ1},
and particular DF integrals appear as (inverse) Fourier transform of $\tau$.

It turns out that the best language study integrability in DF models is that of
Hirota equations in {\it Miwa variables}, which are naturally finite-difference.
These are the equations that describe a special kind of $\tau$-functions
known as ($q$-)Painlev\`e VI $\tau$-function.
Amusingly, they can be rewritten in terms of eight(!) different $\tau$-functions,
which are actually $8$ different shifts of the original KP $\tau$-function.
In this system, one can observe that equations for exactly one half of these functions
are actually related to {\it linear} equations though in a very tricky way.
Not surprisingly, this means that the system is split into two parts:
quadratic integrable and (having linear origin) Ward identities.
Actually, it turns out that they split in exactly equal parts: four and four.
As already mentioned, it would be very interesting to understand, if and how
this result is linked to an alternative unification, that in terms of mysterious
factorization of correlators, single and pair \cite{MMsi,MMd}, nicknamed
superintegrability in \cite{IMM}.
We started to investigate this relation in \cite{MMsi,MMZU,MMd},
but there is still a long way to go.

In DF models, one can restrict the central charge to $c = 1$.
This excludes the $\beta$-deformation
and guarantees that there is no intrinsic breakdown of integrability.
Still, there is a whole variety of deformations
labeled by parameter $t=q$ (in general $t=q^\beta$).
If one is looking at the gauge theory side of the AGT correspondence,
$q\neq 1$ is related to $5d$ gauge theories \cite{5d}, and $q$ encodes the radius of the fifth-dimension.
The limit $q\longrightarrow 1$ corresponds to taking the radius to zero,
when the $5d$ gauge theory reduces to $4d$.
This limit looks ``continuous" at the level of Virasoro-like Ward identities,
when difference operators for $q\neq 1$ become differential for $q=1$.

The aim of the present paper is two-fold.
First of all, we give a review of the above subjects and collect various ideas
scattered through \cite{MMP,MMPq,MMZ}
in a single text.
Second, we explain how different pieces of the construction behave in the non-autonomous limit from $5d$ to $4d$ gauge theory.
We demonstrate that the most important structures survive after taking the limit.
In particular, we follow carefully  the relation ``integrability + string equation = Painlev\`e"
and demonstrate that is consistent with the continuous limit.
Therefore, we make a step in clarifying the relation between the $\tau$-function of the $q$-Painlev\`e VI equations
provided by the $5d$ Nekrasov function and the $\tau$-function of the continuous Painlev\`e VI equations
provided by its $4d$ limit.

Note that the continuous limit is rather simple in terms of the DF matrix model, but is rather complicated in terms of bilinear equations. We illuminate this limiting procedure and outline, how structures such as $q$-Virasoro constraints, bilinear equations and $q$-Painlev\`e equations behave. As an additional result of taking the limit, we obtain another representation of the Painlev\`e $\tau$-function in terms of conformal blocks.

The paper is organized according to this logic.
First, we review general properties of matrix models in Dijkgraaf-Vafa phases (Section 2)
and in Miwa variables (Sections 3).
Then, we start directly from the $q$-deformed case in section 4
and proceed to the description of the continuous limit in section 5.
In section 6, we discuss an important particular case when conformal blocks are degenerate,
which is AGT-related to the pure gauge theory limit.
In this case, the DF model is substituted by the Brez\'in-Gross-Witten (BGW) matrix model \cite{MMShBGW},
and its relation to the Painlev\`e equation becomes especially simple and transparent. This next digression
gives rise to the Painlev\`e III equation.
Conclusion in section 7 briefly summarizes our claims.

\section{Integrability of matrix models in Dijkgraaf-Vafa phases}

\subsection{Hermitian matrix models}

Throughout the paper, we discuss only the Hermitian one matrix model with the partition function given by the integral over $N\times N$ Hermitian matrix $X$:
\be\label{HM}
Z_N^{(\mu)}\{t_k\}={1\over{\rm Vol}_N}\int DX\mu(X) \exp{\left( \sum_{k=1}^{\infty} t_k \Tr X^k \right)}
\ee
Here $DX$ is the Haar measure on Hermitian matrices, $\mu(X)$ is an arbitrary invariant function on them, and ${\rm Vol}_N$ is the volume of the unitary group $U(N)$. The integral is understood as a power series in time variables $t_k$ provided all the moments of the distribution $dX\mu(X)$ are defined. The simplest choice of $\mu(X)$ is the Gaussian distribution:
\be\label{Gauss}
\mu(X)=\exp\Big(-{1\over 2}\Tr X^2\Big)
\ee
One can also integrate out the angular variables in the integral (\ref{HM}), and the remaining integral over eigenvalues $x_i$ of the matrix $X$ is
\be\label{evi}
Z^{(\mu)}_N\{t_k\}={1\over N!}\int_{-\infty}^{\infty}\prod_{i=1}^Ndx_i\mu(x_i)\exp\left(\sum_kt_kx_i^k\right)\Delta^2(x)
\ee
where $\Delta (x)$ is the Vandermonde determinant, $\Delta(x)=\prod_{i>j}(x_i-x_j)$.

\subsection{Dijkgraaf-Vafa phase}

A more tricky case is the choice of a cubic exponential, when one has to choose the pure imaginary coefficient in front of cubic term in order to guarantee convergence properties ($I$ here denotes the imaginary unit):
\be
Z_N^{(3)}\{t_k\}={1\over N!}\int_{-\infty}^{\infty}\prod_{i=1}^Ndx_i\exp\left(Ix_i^3+\sum_kt_kx_i^k\right)\Delta^2(x)
\ee
However, one may consider a more general model eigenvalue model, with the same integrand but with arbitrarily chosen contours. We will always choose the contours going to infinities. Then, in the cubic case, there are two independent ways to choose the contour, these two contours $C_{1,2}$ correspond to two different solutions to the Airy equation. Hence, the partition function of this model is parameterized additionally by two integers $N_1$ and $N_2$ that parameterize the number of contours of each type, and $N_1+N_2=N$ is still the number of $x_i$:
\be\label{DVpf}
Z_{N_1,N_2}^{(3)}\{t_k\}={1\over N_1!N_2!}\int_{C_1}\prod_{i=1}^{N_1}
\int_{C_2}\prod_{i=N_1+1}^{N_1+N_2}dx_i\exp\left(x_i^3+\sum_kt_kx_i^k\right)\Delta^2(x)
\ee
This gives us a typical example of the Dijkgraaf-Vafa phase.

\subsection{Integrability of matrix models}

The partition function (\ref{HM}) has a determinant representation \cite{KMMOZ,versus}:
\be\label{detrep}
Z_N^{(\mu)}\{t_k\}=\det_{1\le i,j\le N}M_{i+j-2}=\det_{1\le i,j\le N}\left({\p M\over\p t_1}\right)^{i+j-2}
\ee
with the moment matrix
\be\label{moment}
M_k:=\int dx\mu(x)x^k\exp\left(\sum_kt_kx^k\right)
\ee
that celebrates the property
\be
{\p M\over\p t_k}=\left({\p M\over\p t_1}\right)^k
\ee
This guarantees that $Z_N^{(\mu)}\{t_k\}$ is a $\tau$-function of the (forced) Toda chain hierarchy and, in particular, of the KP hierarchy w.r.t. time variables $t_k$'s, with $N$ playing the role of zeroth (discrete) time variable \cite{KMMOZ,versus}.

Now, if one considers the DV partition function (\ref{DVpf}), it is no longer a $\tau$-function, and it does not have a determinant representation. However, one may consider the sum
\be
Z_N^{F}\{t_k\}=\sum_{N_1+N_2=N} \xi_1^{N_1}\xi_2^{N_2}Z_{N_1,N_2}^{(3)}\{t_k\}
\ee
This sum can be also represented by the determinant (\ref{detrep}) with the same moment matrix (\ref{moment}) where the integration runs over the formal sum of contours $\xi_1C_1+\xi_2C_2$. Thus, $Z_N^{F}\{t_k\}$ is still a $\tau$-function of the Toda chain hierarchy. Note that, because of the constraint $N_1+N_2=N$, the sum in this formula can be rewritten in the form
\be
Z_N^{F}\{t_k\}=\xi_2^N\sum_{N_1+N_2=N} \xi^{N_1}Z_{N_1,N-N_1}^{(3)}\{t_k\},\ \ \ \ \ \xi:={\xi_1\over\xi_2}
\ee
Hence, up to inessential factor, it is just a discrete Fourier transform of the DV partition function, which gives rise to a $\tau$-function of the integrable hierarchy:
\be\label{FDV}
Z_N^{F}\{t_k\}=\det_{1\le i,j\le N}M_{i+j-2},\ \ \ \ \ \ \ \ \ M_k:=\left(\xi_1\int_{C_1}+\xi_2\int_{C_2}\right)
dx\mu(x)x^k\exp\left(\sum_kt_kx^k\right)
\ee
This is the key observation \cite{MMZ1,MMZ} that is used throughout this paper.

\section{Matrix models in Miwa variables}\label{MMMV}

\subsection{Painlev\`e from integrability and Virasoro constrains}

One of the main features of the matrix model (\ref{HM}) is the existence of an infinite set of Virasoro constraints that it satisfies as a function of times $t_k$: if one chooses measure in the form
\be
\mu(x)=\exp\left(\sum_kT_kx^k\right)
\ee
then the constraints are
\be
L_n Z_N^{(\mu)}\{t_k\} = 0 ,\ \ \ \ \  n \geq -1\nn\\
L_n=\sum_kk(T_k+t_k){\p\over\p t_{k+n}}+\sum_{a=1}^{n-1}{\p^2\over\p t_a\p t_{n-a}}+
2N\frac{\p}{\p t_n}+ N^2\delta_{n,0} + N(T_1+t_1) \delta_{n+1,0}
\ee
These equations reflect the invariance of the integral under any analytic change of integration variables respecting boundary conditions. It is known that reductions of the Virasoro constraints to a small set of time variables often produce Painlev\`e equations, which adds to the long-standing puzzle of the Painlev\`e property of reductions of integrable systems to ODE \cite{Pt}. Let us demonstrate how one can reduce the infinite system of Virasoro constraints for the partition function, which is a $\tau$-function of integrable hierarchy to an ordinary differential equation. When this equation is of the second order, it is often contained in the Painlev\`e list \cite{Plist}.

To begin with, consider the Gaussian model with the measure (\ref{Gauss}), i.e. with $T_k=-{1\over 2}\delta_{k,2}$. Then, the lowest Virasoro constraint $L_{-1}Z_N^{(2)}\{t_k\}=0$, which is called string equation, along with the integrability property guarantees that all the Virasoro constraints are satisfied \cite{UFN3,AMM,Max,MMMs}. Hence, one has the only additional restriction on the $\tau$-function that gives rise to $Z_N^{(2)}\{t_k\}=\tau$, that is, to the string equation
\be
\left(\sum_k(k+1)t_{k+1}{\p\over\p t_{k}}-{\p\over\p t_1}+Nt_1\right)Z_N^{(2)}\{t_k\}=0
\ee
Let us differentiate this equation in $t_1$ and put all times zero but the first three: $t_1=x$, $t_2=y$, $t_3=t$. Then, one obtains from the string equation
\be\label{uy}
2yu_x+3tu_y-u_x=0,\ \ \ \ \ u=\p^2_x\log Z_N^{(2)}\{x,y,t\}
\ee
Differentiating this equation in $x$ and $y$ and inserting the results into the KP equation,
\be\label{KP}
-4u_{xt}+3u_{yy}+6(u^2)_{xx}+u_{xxxx}=0
\ee
one obtains at $y=0$:
\be
-12t^2u_t+u_x-6tu+18t^2(u^2)_x+3t^2u_{xxx}=0
\ee
In order to remove the term $u_t$, one has to use the second Virasoro constraint, $L_{0}Z_N^{(2)}\{t_k\}=0$:
\be
\left(\sum_kkt_{k}{\p\over\p t_{k}}-{\p\over\p t_2}+N^2\right)Z_N^{(2)}\{t_k\}=0
\ee
differentiate it twice in $x$ and use (\ref{uy}) in order to remove $u_y$ and to obtain
\be
3tu_t+2u+xu_x-{u_x\over 3t}=0
\ee
Expressing $u_t$ from this expression, one finally obtains the ordinary differential equation w.r.t. to $x$ with $t$ being just a parameter:
\be
2tu+\left(4xt-{1\over 3}\right)u_x+18t^2(u^2)_x+3t^2u_{xxx}=0
\ee
This is the third order ODE, and, hence, it is not contained in the Painlev\`e list \cite{Plist}.

In order to get a simpler example, one can look at the Kontsevich model \cite{GKM}. The partition function of this model satisfies the Virasoro constraints
\be
\hat{ L}_n^K  Z_K\{t\} = 0, \ \ \ \ n\geq -1\nn\\
\hat { L}_n^K :={1\over 2}\sum_{k} kt_k\frac{\p}{\p t_{k+2n}} + {1\over 4}\sum_{a=1}^{2n-1}\frac{\p^2}{\p t_a\p t_{2n-a}}
+ {t_1^2\over 4}\delta_{n,-1}+{1\over 16}\delta_{n,0}-{\p\over\p t_{2n+3}}
\label{VirK}
\ee
where the sums over $k$ and $a$ run over odd numbers since $ Z_K\{t\}$ does not depend on $t_{2k}$.
Similarly to the Gaussian Hermitian model case, one can consider the case with only $t_1$ and $t_5$ non-zero. However, the ODE that one gets is simpler, since $ Z_K\{t\}$ satisfies the simpler KdV hierarchy, which is the one-dimensional hierarchy. Indeed, from the string equation $\hat{ L}_n^K  Z_K\{t\} = 0$, one obtains at $t_3=0$
\be\label{streq}
5t_5u_t+1=2u_x,\ \ \ \ \ u=\p^2_x\log Z_K
\ee
where we again denote $t_1=x$, $t_3=t$. Now, using the KdV equation, which is the $y$-independent reduction of the KP equation (\ref{KP}),
\be
-4u_{t}+6(u^2)_{x}+u_{xxx}=0
\ee
one reduces (\ref{streq}) to
\be
5t_5u_{xx}+30t_5u^2+4x-8u=const
\ee
This is the second Painlev\`e equation from \cite{Plist} (after a shift of $u$ and rescalings).

\subsection{Hirota equations in Miwa variables}

In the following, however, we are going, to study matrix models in a different parametrization, given by the so called Miwa transformation:
\begin{equation}
    t_k = \dfrac{1}{k} \sum_{a=1}^{\infty} 2\alpha_a z_a^{-k}
\end{equation}
to Miwa variables $z_a$ with multiplicities $\alpha_a$, which have an interpretation of eigenvalues of an external matrix: $t_k = \frac{1}{k} \Tr M^{-k}$ and their multiplicities. After such a transformation, the integral (\ref{evi}) becomes
\begin{equation}\label{MiwaMM}
    Z_{N}\left(z_{a} ; \alpha_{a}\right):=\frac{1}{N !} \int \prod_{i} d x_{i} \mu\left(x_{i}\right) \Delta^{2}(x) \prod_{i, a}\left(1-\frac{x_{i}}{z_{a}}\right)^{2 \alpha_{a}}
\end{equation}
Note that the Miwa factor in this integral,
\be
\prod_a\left(1-\frac{x_{i}}{z_{a}}\right)^{2 \alpha_{a}}=\exp\left(2\sum_a\alpha_a \log\left(1-\frac{x_{i}}{z_{a}}\right)\right)
\ee
can be equally well interpreted as logarithmic additions to the potential of matrix model, which adds more DV phases.
This means that one can govern DV phases by leaving only a finite number of Miwa variables. Any particular choice is in itself a reduction of the infinite set of variables.

Within this setting, we are going to study the Virasoro constraints and the Hirota equations. Even in the reduced space of Miwa variables, it is possible to write down closed expression for (some of) the Virasoro constraints. As for integrability in this case, one can write down the Hirota equations in the Miwa variables \cite{Miwa,GKM}. They becomes bilinear difference equations, for instance,
\begin{equation}\label{hirotamiwa}
\begin{split}
    \left(z_{a}-z_{b}\right) \cdot Z_{N}\left(\alpha_{c}+1 / 2\right) \cdot Z_{N}\left(\alpha_{a}+1 / 2, \alpha_{b}+1 / 2\right)+\left(z_{b}-z_{c}\right) \cdot Z_{N}\left(\alpha_{a}+1 / 2\right) \cdot Z_{N}\left(\alpha_{b}+1 / 2, \alpha_{c}+1 / 2\right)+ \\
+\left(z_{c}-z_{a}\right) \cdot Z_{N}\left(\alpha_{b}+1 / 2\right) \cdot Z_{N}\left(\alpha_{a}+1 / 2, \alpha_{c}+1 / 2\right)=0
\end{split}
\end{equation}
Another example of Hirota bilinear relations, which is of interest for us, involves partition functions with shifts in the zeroth (discrete) discrete Toda time $N$ and, at $z_a=0$, looks like
\begin{equation}\label{Hirota2}
 \begin{array}{r}
z_{b} \cdot Z_{N}\left(\alpha_{c}-1 / 2\right) \cdot Z_{N-1}\left(\alpha_{a}+1 / 2, \alpha_{b}+1 / 2\right)-Z_{N}\left(\alpha_{a}+1 / 2, \alpha_{c}-1 / 2\right) \cdot Z_{N-1}\left(\alpha_{b}+1 / 2\right)- \\
-z_{b} \cdot Z_{N}\left(\alpha_{b}+1 / 2, \alpha_{c}-1 / 2\right) \cdot Z_{N-1}\left(\alpha_{a}+1 / 2\right)=0 \\
\\
z_{c} \cdot Z_{N-1} \cdot Z_{N}\left(\alpha_{a}-1 / 2, \alpha_{b}-1 / 2, \alpha_{c}-1 / 2\right)-Z_{N-1}\left(\alpha_{a}-1 / 2\right) \cdot Z_{N}\left(\alpha_{b}-1 / 2, \alpha_{c}-1 / 2\right)- \\
-z_{c} \cdot Z_{N-1}\left(\alpha_{c}-1 / 2\right) \cdot Z_{N}\left(\alpha_{a}-1 / 2, \alpha_{b}-1 / 2\right)=0
\end{array}
\end{equation}

\subsection{Virasoso constraints in Miwa variables}
The Virasoro constraints in Miwa variables are much less studied, and it is basically unknown how to systematically construct them. We state our current understanding of the problem.

Our observation is that the Virasoro constraints in Miwa parametrization can also be written in a bilinear form. Since we are interested in the DV phase, let us consider the simplest example, a toy example of the Beta-function model,
\begin{equation}
   B_{N}(\alpha_1,\alpha_2) = \int_{0}^{1} \prod_{i=1}^N dx_i \Delta^2(x) \prod_{i=1}^{N} x_i^{2\alpha_1}(1-x_i)^{2\alpha_2}
\end{equation}
We start with the simplest $N=1$ case. Insert a full derivative under the integral to obtain:
\begin{equation}\label{BetaVir}
      0 =  \int_{0}^{1} dx  \dfrac{\partial}{\partial x}  \left[ x^n x^{2\alpha_1}(1-x)^{2\alpha_2} \right] = (2\alpha_1+n) B_{1}\left(\alpha_1+n/2-1/2, \alpha_2 \right) -2\alpha_2 B_{1}\left(\alpha_1, \alpha_2- 1/2 \right)
\end{equation}
This equation definitely depends on the choice of measure, which is here trivial $\mu(x)=1$. On contrary, there is also an equation that is just the first of Hirota equations \eqref{Hirota2} for $N=1$, $z_b=1$ and $Z_0=1$, it looks like
\begin{equation}
    0= \int_{0}^{1} dx \left[ x+(1-x)-1\right]  x^{2\alpha_1}(1-x)^{2\alpha_2} = B_1\left( \alpha_1+1/2, \alpha_2 \right)+B_1\left( \alpha_1, \alpha_2+1/2 \right)-B_1\left( \alpha_1, \alpha_2 \right)
\end{equation}
and does not change with adding a non-trivial measure $\mu(x)$, since it is based on an identical vanishing the integrand. This is the simplest realization of the difference between integrability and Virasoro equations.
\\\\
The $N=1$ measure-dependent equations (\ref{BetaVir}) are linear, however, it seems to be a peculiarity of this simple case, just as this is the case with the Hirota equation. Now let us illustrate the phenomenon that emerges when going to $N=2$. Consider the lowest Virasoro constraint, which would correspond to the following insertion:
\begin{equation}
     0 = \int  d x_1 dx_2  \left( \partial_{x_1} + \partial_{x_2} \right) \left[(x_1-x_2)^2 \prod_{i=1}^{2}  x_i^{2\alpha_1}(1-x_i)^{2\alpha_2}  \right] =  \Phi\left(\alpha_1,\alpha_2\right).
\end{equation}
It is expressed in terms of correlators in the model, however, it can not be represented as an action of some difference operator acting on $\alpha$. However, one can make it bilinear and rearrange the terms in the integrand by renaming the variables to obtain
\begin{equation}\label{betavir1}
    0=\Phi\left( \alpha_1,\alpha_2 \right)\cdot B_1(\alpha_1,\alpha_2) = (2\alpha_1) B_{2}\left( \alpha_1-1/2,\alpha_2\right)B_{1}\left( \alpha_1+1/2,\alpha_2\right) - 2\alpha_2 B_{2}\left( \alpha_1,\alpha_2-1/2\right)B_{1}\left( \alpha_1,\alpha_2+1/2\right).
    \end{equation}
Hence, in order to realize the Virasoro constraints as an operator in Miwa coordinates acting on the partition function, one needs to make them bilinear and use some determinant identities in the integrand. This raises a question whether it is possible to really disentangle the Virasoro and integrability equations in this case. We do not have a clear answer at this point. However, as we see further, the bilinear form is quite natural for ``measure-dependent" relations in the Dotsenko-Fateev matrix model.

The Hirota equation counterpart for the $N=2$ Beta-function model is given by:
 \begin{equation}
 \begin{split}
     B_{2}\left( \alpha_1,\alpha_2\right)B_{1}\left( \alpha_1+\frac{1}{2},\alpha_2+\frac{1}{2}\right) - B_{2} & \left(  \alpha_1+\frac{1}{2},\alpha_2\right)B_{1}\left( \alpha_1,\alpha_2+\frac{1}{2}\right) -
     \\ & - B_{2}\left( \alpha_1,\alpha_2+\frac{1}{2}\right)B_{1}\left( \alpha_1+\frac{1}{2},\alpha_2\right)=0
\end{split}
\end{equation}
The Virasoro equations for the model with three non-zero multiplicities are presented in \cite{MMPq}, here we clarified their origin using a simpler example.

\paragraph{q-deformation.}
The quantum deformation of the matrix model amounts to replacing the integrals with the Jackson integrals,
\begin{equation}
    \int d_q x f(x) = (1-q) \sum_{n=0}^{\infty} q^n f(q^n x)
\end{equation}
and substitutions
\begin{equation}
    \left(1- \dfrac{x_i}{z_a}\right)^{2\alpha_a} \longrightarrow \left(z_a^{-1} x_{i} ; q\right)_{2 \alpha_a}
\end{equation}
where $(x,q)_n$ is the $q$-Pochhammer symbol:
\begin{equation}
    (x,q)_n=\prod_{i=0}^{n-1} (1-q^ix)
\end{equation}
Hence, the integral (\ref{MiwaMM}) becomes
\begin{equation}\label{MiwaMMq}
    Z_{N}\left(z_{a} ; \alpha_{a}\right):=\frac{1}{N !} \int \prod_{i} d_q x_{i} \mu\left(x_{i}\right) \Delta^{2}(x) \prod_{i, a} \left(z_a^{-1} x_{i} ; q\right)_{2 \alpha_a}
\end{equation}
The derivation of the $q$-Virasoro constraints is rather similar to the non-deformed case. Instead of ordinary derivatives, one should use $q$-difference operators
\begin{equation}
D_q^x f(x) =\dfrac{f(qx)-f(x)}{q-1}
\end{equation}
and the property of Jackson integrals
\begin{equation}
    \int d_q x D_q^x f(x)= f(0)
\end{equation}
Let us deal with the example of the $q$-deformed Beta-function model:
\begin{equation}
    B^{(q)}_{N}(\alpha_1,\alpha_2)= \int d_q x_i \Delta^2(x) x_i^{2\alpha_1} (x,q)_{2\alpha_2}
\end{equation}
We use that
\begin{equation}
    \int d_q x_i \sum_{i=1}^{N} D^{x_i}_q \left( \Delta^2(x) x_i^{2\alpha_1} (x,q)_{2\alpha_2} \right) = 0
\end{equation}
and, manipulating with it in a manner similar to the non-deformed case, we obtain a bilinear form of the $q$-Virasoro constraint:
\begin{equation}\label{qVirBeta}
    [2\alpha_1]_q B^{(q)}_{N}\left( \alpha_1-1/2,\alpha_2\right)B^{(q)}_{N-1}\left( \alpha_1+1/2,\alpha_2\right) - [2\alpha_2]_q B^{(q)}_{N}\left( \alpha_1,\alpha_2-1/2\right)B^{(q)}_{N_1}\left( \alpha_1,\alpha_2+1/2\right)=0
\end{equation}
where we used the standard notation for quantum numbers:
\begin{equation}
    [n]_q=\dfrac{q^n-1}{q-1}
\end{equation}
As we see, the $q$-Virasoro constraints in Miwa variables do not differ much with the non-deformed case.

\section{$q$-deformed matrix model and $q$-Painlev\`e equations}

\subsection{DF representation for conformal blocks}

The integral representation of the $c=1$ conformal block of the Virasoro algebra, corresponding to $4d$ gauge theory under the AGT correspondence, is given by the Dotsenko-Fateev representation \cite{MMSh1,AGTmamo},
\be
B(\Delta_i,\Delta,z)=\mathfrak{Z}\cdot Z^{(4d)}_{N}
\ee
\be\label{4dNekr}
Z^{(4d)}_{N}=z^{2 \alpha_{1} \alpha_{2}}(1-z)^{2 \alpha_{2} \alpha_{3}} \cdot \frac{1}{N !} \int \prod_{i} d x_{i} \Delta^{2}(x) \prod x_{i}^{2 \alpha_{1}}\left(1-x_{i}\right)^{2 \alpha_{2}}\left(z-x_{i}\right)^{2 \alpha_{3}}
\ee
where the four external dimensions are parameterized by momenta: $\Delta_i=\alpha_i^2$, $\alpha_4$ is determined from the relation $N+\sum_{i=1}^4\alpha_i=0$, and for the complicated coefficient $\mathfrak{Z}$, see \cite[Eq.(7)-(8)]{MMP}.

(\ref{4dNekr}) is a typical DV type integral similar to (\ref{DVpf}) but the potential is not cubic now, instead it is a sum of three logarithms \cite{MMSh1,MMNek} so that it has also two extremum points, and there are two independent integration contours: $C_1 = [0,z]$ and $C_2 = [1 ,\infty]$. As before, $N_1$ eigenvalues are integrated over $C_1$, and $N_2$ eigenvalues, over $C_2$. The internal dimension of the conformal block (\ref{4dNekr}), $\Delta=\alpha^2$ is determined from the relation
\be\label{cmc}
N_1=\alpha-\alpha_1-\alpha_2,\ \ \ \ \ N_2=-\alpha-\alpha_3-\alpha_4
\ee
This means that, strictly speaking,
the matrix model representations exist only when these two integrality conditions are imposed on the conformal momenta,
while the conformal block at generic values of the external dimensions is obtained by the analytic continuation.
This analytic continuation is immediate for various expansions of the conformal block \cite{MMM}, but not that immediate for the determinant representation (\ref{detrep}) that the matrix model partition function possesses, since it implies a determinant of a matrix of non-integer size.
One possibility to handle this situation is to change a matrix determinant for an infinite-dimensional operator determinant.
This idea was realized on the other side of the AGT story \cite{GIL}. Their approach was actually applicable only for the case, when the conformal momenta satisfy
\be\label{cn1}
\alpha_1\pm\alpha_2+\alpha\notin\mathbb{Z},\ \ \ \ \ \ \ \alpha_1\pm\alpha_2-\alpha\notin\mathbb{Z},\ \ \ \ \ \ \ \alpha_3\pm\alpha_4+\alpha\notin\mathbb{Z},\ \ \ \ \ \ \  \alpha_3\pm\alpha_4-\alpha\notin\mathbb{Z}
\ee
which is complementary to the matrix model restriction on the conformal momenta (\ref{cmc}).
Note that their complicated functional determinants are nothing more than generalizations
of the {\it finite} ones, made from very simple hypergeometric functions,
which arise at the ``integer" locus (\ref{cmc}).

As usual in the DV phase, in order to restore a $\tau$-function of integrable hierarchy, one needs to perform a Fourier transform in the $N_i$ parameters, which can be also understood as a summation in the conformal block internal dimension:
\begin{equation}\label{4dFourier}
    Z_{N}^{(4d)}\left(\xi_{1}, \xi_{2}\right)=\sum_{N_{1}, N_{2}: N_{1}+N_{2}=N} \xi_{1}^{N_{1}} \xi_{2}^{N_{2}} \cdot Z_{N_{1}, N_{2}}^{(4 d)}
\end{equation}
After the Fourier transform from $N_{1,2}$ to $\xi_{1,2}$, (\ref{FDV}) the integration contour is formally given by
\begin{equation}
    \int_{C}=\xi_{1} \int_{C_{1}}+\xi_{2} \int_{C_{2}}
\end{equation}
These integrals correspond to the general matrix model  \eqref{MiwaMM} in the phase where three Miwa variables are non-zero.

Note that, changing variables $x_i\to zx_i$, one transforms the integrals over $C_1$ to those over $C=[0,1]$. Similarly, changing variables $x_i\to x_i^{-1}$, one transforms integrals over $C_2$ to those over C. This allows one to rewrite the integral (\ref{4dNekr}) in the form
\begin{equation}\label{4dN}
   \begin{aligned}
Z_{N_{1}, N_{2}}^{(4d)}=& z^{2 \alpha_{1} \alpha_{2}}(1-z)^{2 \alpha_{2} \alpha_{3}} \cdot \frac{1}{N_{1} ! N_{2} !} \int_0^1 \prod_{i=1}^{N_{1}}\Big( d u_{i} u_{i}^{2 \alpha_{1}}(1-zu_i)^{2 \alpha_{2}}(1-u_i)^{2 \alpha_{3}}\Big) \Delta^{2}(u) \times \\
& \times \int_0^1 \prod_{j=1}^{N_{2}}\Big(d v_{j} v_{j}^{-2 \alpha_{1}-2 \alpha_{2}-2 \alpha_{3}-2 N_{1}-2}(1-v_j)^{2 \alpha_{2}}(1-zv_j)^{2 \alpha_{3}}\Big) \Delta^{2}(v) \times \prod_{i=1}^{N_{1}} \prod_{j=1}^{N_{2}}\left(1-z u_{i} v_{j}\right)^{2}
\end{aligned}
\end{equation}

\bigskip

Similarly, in the $5d$ gauge theory case, the starting point is the $q$-Virasoro conformal block, AGT dual to $5d$ gauge theory, and its realization in terms of matrix model ($q$-Selberg) integrals is \cite{5dmamo}
\begin{equation}\label{5dNekr}
   \begin{aligned}
Z_{N_{1}, N_{2}}^{(5 d)}=& z^{2 \alpha_{1} \alpha_{2}}(z ; q)_{2 \alpha_{2} \alpha_{3}} \cdot \frac{1}{N_{1} ! N_{2} !} \int \prod_{i=1}^{N_{1}}\left(z^{2 \alpha_{1}+2 \alpha_{2}+N_{1}} d_{q} u_{i} u_{i}^{2 \alpha_{1}}\left(u_{i} ; q\right)_{2 \alpha_{2}}\left(z u_{i} ; q\right)_{2 \alpha_{3}}\right) \Delta^{2}(u) \times \\
& \times \int \prod_{j=1}^{N_{2}}\left(d_{q} v_{j} v_{j}^{-2 \alpha_{1}-2 \alpha_{2}-2 \alpha_{3}-2 N_{1}-2}\left(z v_{j} ; q\right)_{2 \alpha_{2}}\left(v_{j} ; q\right)_{2 \alpha_{3}}\right) \Delta^{2}(v) \times \prod_{i=1}^{N_{1}} \prod_{j=1}^{N_{2}}\left(1-z u_{i} v_{j}\right)^{2}
\end{aligned}
\end{equation}
In order to restore a $\tau$-function of integrable hierarchy, one again needs a Fourier transform in the $N_i$ parameters, which can be also understood as a summation in the conformal blocks internal dimension:
\begin{equation}\label{5dFourier}
    Z_{N}^{(5 d)}\left(\xi_{1}, \xi_{2}\right)=\sum_{N_{1}, N_{2}: N_{1}+N_{2}=N} \xi_{1}^{N_{1}} \xi_{2}^{N_{2}} \cdot Z_{N_{1}, N_{2}}^{(5 d)}
\end{equation}

\subsection{Determinant representation for $q$-Virasoro conformal block}

Now we first study the $q$-deformed case, since, even though the formulas are more involved in this case, they are actually better controllable.

One can write down the determinant representation (\ref{detrep}) for the Fourier transform of the partition functions (which provides integrability):
\begin{equation}
    Z_{N}^{(5 d)}\left(\xi_{1}, \xi_{2}\right)=z^{2 \alpha_{1} \alpha_{2}}(z ; q)_{2 \alpha_{2} \alpha_{3}} \cdot \det_{1 \leq i, j \leq N} M_{i+j-2}
\end{equation}
where the moment matrix (\ref{moment}) is, in this case,
\begin{equation}
\begin{split}
M_k=&\xi_{1} z^{2 \alpha_{12}+k+1} \int d_{q} u u^{2 \alpha_{1}+k}(u ; q)_{2 \alpha_{2}}(q^{-1}z u ; q)^{-1}_{-2 \alpha_{3}}+\\&+\xi_{2} \, q^{-\left(2 \alpha_{1}+1\right)\left(2 \alpha_{23}+1\right)}\int d_{q} v v^{-2 \alpha_{1}-2 \alpha_{2}-2 \alpha_{3}-2-k}(q^{-1}z v ; q)^{-1}_{-2 \alpha_{2}}(v ; q)_{2 \alpha_{3}}= \\
=&\xi_{1} \cdot z^{2 \alpha_{12}+k+1} \cdot \mathfrak{B}_{q}\left(2 \alpha_{1}+k+1,2 \alpha_{2}+1\right) \, { }_{2} \phi_{1}\left(q^{-2 \alpha_{3}}, q^{2 \alpha_{1}+k+1} ; q^{2 \alpha_{12}+k+2} ; q, z\right)+ \\
&+\xi_{2} \cdot q^{-\left(2 \alpha_{1}+1\right)\left(2 \alpha_{23}+1\right)} \cdot \mathfrak{B}_{q}\left(-2 \alpha_{123}-k-1,2 \alpha_{3}+1\right){ }_{2} \phi_{1}\left( q^{-2 \alpha_{2}}, q^{-2 \alpha_{123}-k-1} ; q^{-2 \alpha_{12}-k} ; q, z\right)
\end{split}
\end{equation}
Here the $q$-Beta functions are given by:
\begin{equation}
\mathfrak{B}_{q}(\alpha, \beta)=\int_{0}^{1} d_{q} x x^{\alpha-1}(x ; q)_{\beta-1}=\frac{\Gamma_{q}(\alpha) \Gamma_{q}(\beta)}{\Gamma_{q}(\alpha+\beta)}\end{equation}
with $\Gamma_{q}(\alpha) $ being the $q$-$\Gamma$-function, and
$\phantom{}_2\phi_1$ is the Heine basic hypergeometric series:
\begin{equation}
    { }_{2} \phi_{1}(a, b ; c ; q, z):=\sum_{n=0}^{\infty} \frac{(a ; q)_{n}(b ; q)_{n}}{(c ; q)_{n}(q ; q)_{n}} z^{n}
\end{equation}
For the sake of simplicity, some of the formulas below will be presented in the $\xi_2=0$ case.

\subsection{q-deformed Painlev\`e VI equation}
The $q$-Painlev\`e VI equations are given by \cite{JS}:
\be\label{qPVI1}
\frac{w_{1}(z) w_{1}(q z)}{a_{3} a_{4}}=\frac{\left(w_{2}(q z)-b_{1} z\right)\left(w_{2}(q z)-b_{2} z\right)}{\left(w_{2}(q z)-b_{3}\right)\left(w_{2}(q z)-b_{4}\right)} \\
\frac{w_{2}(z) w_{2}(q z)}{b_{3} b_{4}}=\frac{\left(w_{1}(z)-a_{1} z\right)\left(w_{1}(z)-a_{2} z\right)}{\left(w_{1}(z)-a_{3}\right)\left(w_{1}(z)-a_{4}\right)}\label{qPVI2}
\ee
The consistency condition for these equations requires the relation
\begin{equation}
\frac{b_{1} b_{2}}{b_{3} b_{4}}=q \frac{a_{1} a_{2}}{a_{3} a_{4}}
\end{equation}
Rescaling the variables $z, w_1 $ and $w_2$ and using the consistency condition, one can reduce the number of independent parameters to four. Throughout the paper, we use the parametrization dictated by association of the solution with the conformal block:
\begin{equation}\label{conformalparameters}
    \begin{aligned}
&\alpha_1+\alpha_2+\alpha_3+\alpha_4+N=0, \quad a_{1}=q, \quad a_{2}=q^{1-N-2 \alpha_{3}}, \quad a_{3}=q^{2-N}, \quad a_{4}=q^{2 \alpha_{2}+1} \\
&b_{1}=q^{-2 \alpha_{2}+1}, \quad b_{2}=q^{2 \alpha_{1}+2 \alpha_{3}+N+1}, \quad b_{3}=q^{2 \alpha_{3}+1}, \quad b_{4}=q^{2 \alpha_{1}+2 \alpha_{3}+N+1}
\end{aligned}
\end{equation}
Note that there are exactly four free parameters $q^{\alpha_i}, \  i=1,2,3,4$.
\\\\
From equation \eqref{qPVI1}, one can find $w_2(qz)$ as a function of $w_{1}(z) w_{1}(q z)$ solving the quadratic equation. After rescaling the variable $z \rightarrow q^{-1}z$,
\be \label{qPVI3}
\frac{w_{1}(q^{-1}z) w_{1}(z)}{a_{3} a_{4}}=\frac{\left(w_{2}(z)-b_{1}q^{-1} z\right)
\left(w_{2}(z)-b_{2}q^{-1} z\right)}{\left(w_{2}(z)-b_{3}\right)\left(w_{2}(z)-b_{4}\right)}
\ee
one can similarly find $w_2(z)$ as a function of $w_{1}(q^{-1}z) w_{1}(z)$. Inserting these $w_2(z)$ and $w_2(qz)$ into equation \eqref{qPVI2}, one finds an equation expressing $w_1(qz)$ as a double-valued function of $w_1(z)$ and $w_1(q^{-1}z)$, which represents a difference counterpart of the second order differential equation. The difference Painlev\`e equation shares various properties with the continuous equations, such as a certain analogue of the Painlev\`e property and a rich group of B{\"a}cklund transformations.

\subsection{8 equations}

The system of $q$-deformed Painlev\`e equations can be represented  in terms of 8 bilinear equations for the $\tau$-functions. One of the ways to obtain these equations is to study a discrete counterpart of the Painlev\`e property, i.e. moving of singularities of the equations. It is called singularity confinement criterion \cite{RGH}.\\
The discrete evolution starts with some initial condition $w_1(z)$. Suppose that $w_1(z)$ reaches the pole $a_3$ in the course of evolution so that $w(q^{-1}z)$ is not at the pole yet. We assume generic initial data, which implies all the zeroes and poles are reached at different values of $z$. Now, at the pole $w_1(z)=a_3$, $w_2(z)$ is finite: otherwise, equation (\ref{qPVI3}) leads to $w_1(q^{-1}z)=a_4$, i.e. to the pole value. As soon as $w_2(z)$ is finite at $w_1(z)=a_3$, $w_2(qz)=\infty$, and, as follows from (\ref{qPVI1}), $w_1(qz)=a_4$. This means that, at the next step in evolution, $w_1$ again gets to pole. However, one cannot determine the result of the evolution at this next step: the equations are satisfied by a generic $w_2(q^2z)$. This is just the singularity called confined. A reversed singularity pattern appears if one starts with $w_1(z)=a_4$ instead. Moreover, a similar consideration is applicable to zeroes of these equations. There are in total 8 such patterns, four of which are associated with singularities, and four others, with zeros. From these singularity patterns, one can deduce the bilinear representation of the $q$-Painlev\`e system \cite{TM}.

In order to describe it, we define 8 functions $\tau_i(z)$ implicitly depending on the parameters of the $q$-Painlev\`e equation. These $\tau$-functions are functions on the weight lattice of $D^{(1)}_5$  with symmetries under the affine Weyl group \cite{TM} 9see also \cite{Tak}), and they are related to the functions $w_{1,2}(z)$ by the relations
\begin{equation}\label{qPw}
\begin{split}
w_{1}(z)= z \, \frac{\tau_{1}(q z) \tau_{2}(z)}{\tau_{3}(q z) \tau_{4}(z)} \\
w_{2}(z)= z \dfrac{b_1 a_4}{a_2 a_3}  \cdot \frac{\tau_{5}(z) \tau_{6}(z)}{\tau_{7}(z) \tau_{8}(z)}
\end{split}
\end{equation}

The described singularities are encoded in quadratic relations between the $\tau$-functions equivalent to the $q$-Painlev\`e system \cite{N,TM,JNS}. In the parametrisation \eqref{conformalparameters}, they take the form:
\begin{equation}
\begin{split}\label{8qint}
&z q^{2 N-2} \tau_{1} \tau_{2}-q^{2 \alpha_{2}} \tau_{3} \tau_{4}-\tau_{7} \tau_{8}=0 \\
&\tau_{1} \tau_{2}-q^{1-2 \alpha_{3}-2 N} \tau_{3} \tau_{4}-\tau_{5} \tau_{6}=0 \\
&\bar{\tau}_{1} \tau_{2}-q^{1-N} \bar{\tau}_{3} \tau_{4}-q^{2 N-2 \alpha_{12}-2} \tau_{5} \bar{\tau}_{6}=0 \\
&z q^{N-1} \bar{\tau}_{1} \tau_{2}-q^{2 \alpha_{2}} \bar{\tau}_{3} \tau_{4}-\bar{\tau}_{7} \tau_{8}=0 \\
\end{split}
\end{equation}
\begin{equation}
\begin{split}\label{8qvir}
&z \bar{\tau}_{1} \tau_{2}-q^{2-2 N} \bar{\tau}_{3} \tau_{4}-q^{-2 \alpha_{2}} \tau_{7} \bar{\tau}_{8}=0 \\
&\bar{\tau}_{1} \tau_{2}-q^{1-2 N-2 \alpha_{3}} \bar{\tau}_{3} \tau_{4}-\bar{\tau}_{5} \tau_{6}=0
\\
&\bar{\tau}_{1} \underline{\tau}_{2}-q^{1-N} \bar{\tau}_{3} \underline{\tau}_{4}-q^{2 N-2-2 \alpha_{12}} \tau_{5} \tau_{6}=0 \\
&z \bar{\tau}_{1} \underline{\tau}_{2}-q^{2-N} \bar{\tau}_{3} \underline{\tau}_{4}-q^{N-2 \alpha_{2}} \tau_{7} \tau_{8}=0
\end{split}
\end{equation}
where we denoted $\bar\tau=\tau(qz)$ and $\underline{\tau}=\tau(q^{-1}z)$.
In fact, these eight $\tau$-functions can be expressed through a single $\tau$-function $\tau(\alpha_1,\alpha_2,\alpha_3,\alpha_4;z)$, which is a function of $z$ and four parameters $\alpha_i , i = 1 \ldots 4$ with different $\tau$-functions corresponding to certain shifts in the parameter space:
\begin{equation}
  \begin{array}{cc}
\tau_{1}\left(\alpha_{1}, \alpha_{2}, \alpha_{3}, \alpha_{4} ; z\right)=\tau\left(\alpha_{1}+\frac{1}{2}, \alpha_{2}, \alpha_{3}+\frac{1}{2}, \alpha_{4} ; z\right) & \tau_{2}\left(\alpha_{1}, \alpha_{2}, \alpha_{3}, \alpha_{4} ; z\right)=\tau\left(\alpha_{1}, \alpha_{2}-\frac{1}{2}, \alpha_{3}, \alpha_{4}+\frac{1}{2} ; z\right) \\
\tau_{3}\left(\alpha_{1}, \alpha_{2}, \alpha_{3}, \alpha_{4} ; z\right)=\tau\left(\alpha_{1}, \alpha_{2}, \alpha_{3}+\frac{1}{2}, \alpha_{4}+\frac{1}{2} ; z\right) & \tau_{4}\left(\alpha_{1}, \alpha_{2}, \alpha_{3}, \alpha_{4} ; z\right)=\tau\left(\alpha_{1}+\frac{1}{2}, \alpha_{2}-\frac{1}{2}, \alpha_{3}, \alpha_{4} ; z\right) \\
\tau_{5}\left(\alpha_{1}, \alpha_{2}, \alpha_{3}, \alpha_{4} ; z\right)=\tau\left(\alpha_{1}+\frac{1}{2}, \alpha_{2}, \alpha_{3}, \alpha_{4}+\frac{1}{2} ; z\right) & \tau_{6}\left(\alpha_{1}, \alpha_{2}, \alpha_{3}, \alpha_{4} ; z\right)=\tau\left(\alpha_{1}, \alpha_{2}-\frac{1}{2}, \alpha_{3}+\frac{1}{2}, \alpha_{4} ; z\right) \\
\tau_{7}\left(\alpha_{1}, \alpha_{2}, \alpha_{3}, \alpha_{4} ; z\right)=\tau\left(\alpha_{1}+\frac{1}{2}, \alpha_{2}-\frac{1}{2}, \alpha_{3}+\frac{1}{2}, \alpha_{4}+\frac{1}{2} ; z\right) & \tau_{8}\left(\alpha_{1}, \alpha_{2}, \alpha_{3}, \alpha_{4} ; z\right)=\tau\left(\alpha_{1}, \alpha_{2}, \alpha_{3}, \alpha_{4} ; z\right)
\end{array}
\end{equation}
The parameters are once again chosen to facilitate further representation of the $\tau$-functions in terms of conformal blocks and the logarithmic model. Equations (\ref{8qint},\ref{8qvir}) turn into bilinear difference equations for the function $\tau(\alpha_1,\alpha_2,\alpha_3,\alpha_4;z)$. These equations resemble the bilinear difference equations that we reviewed in section \ref{MMMV}. We discuss these equations from this point of view for the rest of this section and later study their continuous limit.

From the discussion above, one should expect that, since the $w_2(z)$ function can be eliminated, the four $\tau$-functions $\tau_{5,6,7,8}$ are, in a sense, also auxiliary. Moreover, one can express $w_2(z)$ from the first four $\tau$-functions, similarly to $w_1(z)$: to this end, one can just use for this the first two equations of (\ref{8qint}), or the last two equations of (\ref{8qvir}). However, eliminating the last four $\tau$-functions would make the structure of remaining four equations far more involved and non-transparent.

\paragraph{Conformal block solves the q-Painlev\`e equation.}
Having described the bilinear form of the $q$-Painlev\`e equation, we are now ready to describe how the partition function of the logarithmic q-deformed matrix model provides the $q$-Painlev\`e $\tau$-function. Recall the relation between the conformal parameters and the matrix integral:
\begin{equation}
    \alpha_1+\alpha_2+\alpha_3+\alpha_4=-N.
\end{equation} Then we can identify the two functions as follows:
\begin{equation}
 \tau\left(\alpha_{1}, \alpha_{2}, \alpha_{3}, \alpha_{4} ; z\right)=  z^{-2 \alpha_{1} \alpha_{2}}(z ; q)^{-1}_{2 \alpha_{2} \alpha_{3}} Z^{(5d)}_N\left(\alpha_{1}, \alpha_{2}, \alpha_{3}  \right)
\end{equation}
This statement was first explored on the conformal side in \cite{JNS}. In terms of the matrix model, the claim amounts to the partition function that satisfies relations (\ref{8qint},\ref{8qvir})
With the relation between $\alpha_4$ and $N$, the left column of the $\tau$-functions with odd indices corresponds to shift $N \rightarrow N-1$, hence the relations are in terms of quadratic combinations of the form $Z_N Z_{N-1}$.
In other words, solutions to the $q$-Painlev\`e VI equations are provided by the following ratios of partition functions:
\begin{equation}
    \begin{split}
        w_1(z)&= q^{N} z \dfrac{Z^{(5d)}_N\left(\alpha_1+\frac{1}{2} ,qz\right)Z^{(5d)}_{N-1}\left(\alpha_2-\frac{1}{2} ,z\right)}{Z^{(5d)}_N\left(\alpha_3+\frac{1}{2}  ,qz\right)Z^{(5d)}_{N-1}\left(\alpha_1+\frac{1}{2} ,\alpha_2-\frac{1}{2}  ,z\right)} \\ w_2(z)&= q^{2 \alpha_{3}+2 N-1} z  \dfrac{Z^{(5d)}_N\left(\alpha_1+\frac{1}{2}  ,z\right)Z^{(5d)}_{N-1}\left(\alpha_2-\frac{1}{2},\alpha_3+\frac{1}{2}   ,z\right)}{Z^{(5d)}_N\left(\alpha_1+\frac{1}{2},\alpha_2-\frac{1}{2} ,\alpha_3+\frac{1}{2}   ,z\right)Z^{(5d)}_{N-1}\left( z\right)}
    \end{split}
\end{equation}
where we have listed only the variables that are shifted.
\\\\
The bilinear equations (\ref{8qvir},\ref{8qint}) are our main object of study. We treat them from the perspective of the matrix model representation of Nekrasov functions.  From the matrix model perspective, these equations have a distinguished meaning. In fact, they split into two sets of equations. By an explicit check, one can notice that equations \eqref{8qvir} do not depend on the matrix model measure, which means they represent the Hirota equations, responsible for integrability of partition functions.

To see this, we look at the eigenvalue representation \eqref{MiwaMMq}, and notice that it can be treated as a special case of the integral \eqref{MiwaMM}, with a special choice of Miwa variables:
\begin{equation}\label{qmiwa}
    \left(0,2 \alpha_{1}\right), \quad\left(z q^{-i}, 1\right), \quad i=0, \ldots, 2 \alpha_{2}-1, \quad\left(q^{-i}, 1\right), \quad i=0, \ldots, 2 \alpha_{3}-1
\end{equation}
and the integral substituted by the Jackson integral. The difference between the ordinary integration and the Jackson one is inessential for the Hirota equations, because, as we saw, their origins are identities of the integrand. In other words, one may think of the Jackson integral as a specific contour integral of a function with simple poles at points \eqref{qmiwa}, while the choice of integration contour does not affect the Hirota equations.

However, the other equations \eqref{8qvir} are measure dependent. For example, one can easily check that changing the integration measure in \eqref{5dNekr} as $\prod\limits_{i} d_q x_i \rightarrow \prod\limits_{i} x_i d_q x_i$ simply shifts $\alpha_1 \rightarrow \alpha_1+1/2$ in the $\tau$-functions. At the same time, we keep the parameter $\alpha_1$ in the coefficients of equations \eqref{8qvir}, \eqref{8qint} intact. Then, equation \eqref{8qint} still holds, while \eqref{8qvir} do not.

If we started with the Hirota equations, we could conclude that the $\tau$-function can be represented by an eigenvalue integral with an arbitrary measure. This would imply that the space of $\tau$-functions is as large as the space of solutions of the whole (forced) Toda chain hierarchy. The role of the other four equations \eqref{8qvir} is to fix the specific logarithmic measure: they represent a reduction of the KP hierarchy, specifically to solutions of the $q$-Painlev\`e equations.

In this sense, they are nothing but the Virasoro constraints. As we have discussed above, the bilinear form of the Virasoro constraints is a feature of the Miwa parametrization. Hence the system \eqref{8qvir}+\eqref{8qint} plays a role of the usual ``integrability+string equation" pair in matrix models. We see that, in this case, the structure is not as transparent as usual, since the reduction requires rather complicated combinations of the Virasoro constraints. As we will see below, it seems this complication is not an issue of the $q$-deformation, in fact the $q$-deformed case seems to be the clearest one. It most likely that the lack of understanding of the Virasoro constraints in the Miwa parametrization is the source of the current problems.
\\\\
Let us demonstrate how this works in the $N=1$ case, when the equations are linear. In this case, the measure dependent equations simplify, and also become a corollary of integrability, and only the first equation of \eqref{8qvir} remains independent. Hence, we take the first equations from \eqref{8qint} and \eqref{8qvir}:
\begin{equation}
\begin{split}
    z \tau_2 - q^{2\alpha_2} \tau_4 -\tau_8&=0
\\
    z \tau_2 -  \tau_4 - q^{-2\alpha_2}\taub_8&=0
\end{split}
\end{equation}
Despite the striking similarity, we can easily see a different nature of these equation. Indeed, let us rewrite them in terms of the partition function, and make an overall shift $\alpha_2 \rightarrow \alpha_2+1/2$ for convenience:
\begin{equation}
\begin{split}
        z Z^{(5d)}_{1} \left(z \right) - q^{2\alpha_2+1} Z^{(5d)}_{1} \left(\alpha_1+1/2, z \right) -  Z^{(5d)}_{1} \left(\alpha_2+1/2,  z\right) =0
        \\
        z Z^{(5d)}_{1} \left(\alpha_2 ,z \right) - Z^{(5d)}_{1} \left(\alpha_1+1/2,\alpha_2, z \right) -q^{-2\alpha_2-1}   Z^{(5d)}_{1} \left(\alpha_2+1/2, q z\right) =0
\end{split}
\end{equation}
where we have written explicitly only the shifted multiplicities. Let us represent these equations in terms of the expectation value in the integral with intact parameters:
\begin{equation}
\begin{split}
        \left\langle z - q^{2\alpha_2+1} z x - z\left(1-q^{2\alpha_2+1}x \right) \right\rangle_{\alpha_1,\alpha_2,\alpha_3} & =0
\\
  \left\langle z - z x \right\rangle_{\alpha_1,\alpha_2,\alpha_3} - \left\langle z\left(q^{2\alpha_1}(1-zq^{-1}x \right) \right\rangle_{\alpha_1,\alpha_2+1/2,\alpha_3-1/2} & =0
\end{split}\end{equation}
The first identity is a trivial identity that holds without any integration, hence it does not require fixing any specific measure. The second identity is, however, a consequence of the full $q$-derivative insertion:
\begin{equation}
  0 =  \int d_q x D_q \left( x^{2\alpha_1} \left(x;q)_{2\alpha_2+1}\right)\left(z x;q \right)^{-1}_{-2\alpha_3}\right)
\end{equation}
Therefore, it is sensitive to changes in the measure. In this simple case, it is simple to identity the full-derivative insertion, explicitly showing how it is related to the ``basic" Virasoro constraint.
For generic $N$, the equations \eqref{8qvir} are analogues of the Virasoro equations \eqref{qVirBeta} for the generalized $q$-Beta function, but now with 3 non-zero multiplicities.

Note that, in this case, the Virasoro constraints are difference equations in $\alpha$ but also in $z$, which puts them more on an equal footing. This is a natural property of $q$-hypergeometric functions, which are components of the conformal block. For example, recall the Heine symmetry of the basic hypergeometric series:
\begin{equation}
    { }_{2} \phi_{1}(a, b ; c ; q, z)=\frac{(b ; q)_{\infty}(a z ; q)_{\infty}}{(c ; q)_{\infty}(z ; q)_{\infty}}{ }_{2} \phi_{1}(c / b, z ; a z ; q, b)
\end{equation}
which mixes $a,b,c$ and $z$.

\section{$q$-Painlev\`e to Painlev\`e VI ($5d$ to $4d$) limit}
Having described the different components in $q$-Painlev\`e theory, we proceed to describe the continuous limit of all of them.

\subsection{Conformal block}

The limit $q$-Virasoro to Virasoro for the conformal blocks themselves is just straightforward.
As it follows from (\ref{4dN}), the conformal block has the determinant representation
\begin{equation}
Z_{N}\left(\xi_{1}, \xi_{2}\right)=z^{2 \alpha_{1} \alpha_{2}}(1-z)^{2 \alpha_{2} \alpha_{3}} \cdot \det_{1 \leq i, j \leq N} M_{i+j-2}
\end{equation}
where
\begin{equation}
    \begin{split}
M_k=\xi_{1} \int_{0}^{z} x^{2 \alpha_{1}+k}(1-x)^{2 \alpha_{2}}(z-x)^{2 \alpha_{3}} d x+\xi_{2} \int_{1}^{\infty} x^{2 \alpha_{1}+k}(1-x)^{2 \alpha_{2}}(z-x)^{2 \alpha_{3}} d x= \\
=\xi_{1} z^{2 \alpha_{12}+k+1} \mathfrak{B}\left(2 \alpha_{1}+k+1,2 \alpha_{2}+1\right) \quad{ }_{2} F_{1}\left(-2 \alpha_{3}, 2 \alpha_{1}+k+1 ; 2 \alpha_{12}+k+2 ; z\right)+ \\
+\xi_{2} \mathfrak{B}\left(-2 \alpha_{123}-k-1,2 \alpha_{3}+1\right) \quad{ }_{2} F_{1}\left(-2 \alpha_{123}-k-1,-2 \alpha_{2} ;-2 \alpha_{12}-k ; z\right)
\end{split}
\end{equation}
and $\mathfrak{B}$, ${ }_{2} F_{1}$ are the usual Beta-function and the hypergeometric function accordingly.

The Virasoro constraints \eqref{betavir1} are appropriately modified, see, for example \cite{MMPq}. They involve additional $z$ dependent terms and shifts of the $\alpha_3$ variable. Note that, in the continuous limit, the multiplicity values $\alpha_i$ and $z$ are now less symmetric. For instance, the Virasoro constraints are difference equations in $\alpha$ and differential ones in $z$.

\subsection{Painlev\`e VI equation}

The famous result of \cite{Gamayun:2013auu} states that the conformal block solves the Painlev\`e VI equation. Unlike the $q$-deformed case, there are various different but equivalent forms of the equation that appear in the discussion. Let us start by collecting various forms of the equation relevant for our discussion.

The standard form for the Painlev\`e equations is the second order equation for a function $w(z)$, resembling a Newton equation, hence we call it Newton form:
\begin{equation}\label{PVInewt}
\begin{split}
\frac{d^{2} w}{d z^{2}}=\frac{1}{2} &\left(\frac{1}{w}+\frac{1}{w-1}+\frac{1}{w-z}\right)\left(\frac{d w}{d z}\right)^{2}-\left(\frac{1}{z}+\frac{1}{z-1}+\frac{1}{w-z}\right) \frac{d w}{d z}+\\
&+\frac{2 w(w-1)(w-z)}{z^{2}(z-1)^{2}}\left(\left(\theta_\infty+\frac{1}{2}\right)^2-\frac{\theta_0^2 z}{w^{2}}+\frac{\theta_1^2(z-1)}{(w-1)^{2}}+\frac{\left(\frac{1}{4}-\theta_z^2\right) z(z-1)}{(w-z)^{2}}\right)
\end{split}
\end{equation}
It is also well-known that, for this equation, one can introduce a Hamiltonian $ H_{VI}$ and momentum $p(z)$, and obtain the Hamiltonian form:
\begin{equation}\label{PVIhamilt}
\begin{split}
\dfrac{d w}{d z}=-\dfrac{w(w-1)(w-z)}{z(z-1)}\left(2 p-\dfrac{2 \theta_0}{w}-2\dfrac{2\theta_1}{w-1}-\dfrac{2\theta_z-1}{w-z}\right)=\dfrac{\partial H_{VI}}{\partial p}
\end{split}
\end{equation}
Finally, if one starts with the problem of isomonodromic deformations of the Shlesinger systems, one obtains the Painlev\`e equation in the so-called $\sigma$-form:
\begin{equation}\label{PVIsigma}
    \left(z(z-1) \sigma^{\prime \prime}\right)^{2}=-2 \operatorname{det}\left(\begin{array}{ccc}
2 \theta_{1}^{2} & z \sigma^{\prime}-\sigma & \sigma^{\prime}+\theta_{1}^{2}+\theta_{2}^{2}+\theta_{3}^{2}-\theta_{4}^{2} \\
z \sigma^{\prime}-\sigma & 2 \theta_{2}^{2} & (z-1) \sigma^{\prime}-\sigma \\
\sigma^{\prime}+\theta_{1}^{2}+\theta_{2}^{2}+\theta_{3}^{2}-\theta_{4}^{2} & (z-1) \sigma^{\prime}-\sigma & 2 \theta_{3}^{2}
\end{array}\right)
\end{equation}

The relation between these presentations is cumbersome but known and given by the following relations \cite{Gamayun:2013auu}. The variable $\sigma$ is related to the Hamiltonian variables as
\begin{equation}
    \begin{split}
\sigma=&z(z-1) H_{\mathrm{VI}}-w(w-1) p+\left(\theta_{0}+\theta_{z}+\theta_{1}+\theta_{\infty}\right) w - \\
&-\left(\theta_{0}+\theta_{1}\right)^{2} z+\frac{\theta_{1}^{2}+\theta_{\infty}^{2}-\theta_{0}^{2}-\theta_{z}^{2}-4 \theta_{0} \theta_{z}}{2}
\end{split}
\end{equation}
and to the Newtonian form:
\begin{equation}\label{sigmatonewton}
    \begin{split}
 \frac{1}{w-z}+\frac{1}{2}\left(\frac{1}{z}+\frac{1}{z-1}\right)&=\\
=& \frac{2 \theta_{\infty} z(z-1) \sigma^{\prime \prime}+\left(\sigma^{\prime}+\theta_{z}^{2}-\theta_{\infty}^{2}\right)\left((2 z-1) \sigma^{\prime}-2 \sigma+\theta_{0}^{2}-\theta_{1}^{2}\right)+4 \theta_{\infty}^{2}\left(\theta_{0}^{2}-\theta_{1}^{2}\right)}{2 z(z-1)\left(\sigma^{\prime}+\left(\theta_{t}-\theta_{\infty}\right)^{2}\right)\left(\sigma^{\prime}+\left(\theta_{t}+\theta_{\infty}\right)^{2}\right)}
    \end{split}
\end{equation}

\paragraph{Conformal block solves the Painlev\`e VI equation.}
There is also another, bilinear form of the Painlev\`e equation: if one makes the change of variables to a Painlev\`e $\tau$-function,
\begin{equation}
    \sigma(z)=z(z-1)\dfrac{d \log \tau_P}{dz}
\end{equation}
the Painlev\`e VI equation can be written in the form  \cite{Bershtein:2014yia}:
\begin{equation}
    D^{VI} \tau_P\cdot\tau_P = 0
\end{equation}
\begin{equation}\label{BE}
    D^{V I}=-\frac{1}{2}(1-z)^{3} D_{\log z}^{4} + \ldots
\end{equation}
where $D$ is the standard Hirota derivative operator defined as
\begin{equation}
D_h^k f(x)\cdot g(x)=\left.\left({\p^k\over \p h^k}f(x+h)g(x-h)\right)\right|_{h=0}
\end{equation}
In such a form, it is a rather cumbersome equation of order 4 in the Hirota operators.

Now we can formulate the relation between the $4d$ matrix model partition function (\ref{4dNekr}) and the Painlev\`e VI equation: the partition function is just the Painlev\`e $\tau$-function \cite{MMP}
\begin{equation}
Z^{(4d)}_N=\tau_P
\end{equation}
or
\begin{equation}\label{PVIsigmasolution}
    \sigma(z)=z(z-1)\dfrac{d \log \log Z^{(4d)}_N}{dz}
\end{equation}
with parameters of the equation related to the operator dimensions by
\begin{equation}\label{parameters}
    \theta_0 = \alpha_1\, ,\theta_z = \alpha_2\, ,\theta_1 = \alpha_4\, ,\theta_\infty = \alpha_4\,
\end{equation}
There are different ways to see that \eqref{PVIsigmasolution} satisfies \eqref{PVIsigma}. By setting $N=0$, one checks that the prefactor $Z_0=z^{2\alpha_1 \alpha_2}(1-z)^{2\alpha_2 \alpha_3}$ satisfies the equation. For nonzero $N$, the partition functions is given by the non-perturbative prefactor times a power series in $z$, and one checks order by order that the coefficients of the $z$ expansion of $Z_N$ satisfy the equation.
\\\\
Looking at the bilinear equation (\ref{BE}) from a matrix model perspective, we see that this equation does depend on changing the measure, hence we should treat it some combination of Virasoro constraints. In such formulation, it may seem that this is a single bilinear equation equivalent to the Painlev\`e equation. However, it turns out not to be the case.
To make a connection with the next section, we note that generally the Painlev\`e equations can be written as bilinear equations in many different ways. The one described above uses one single $\tau$-function. However, one could introduce multiple $\tau$-functions related by B{\"a}cklund transformation \cite{Bershtein:2016uov}, just as in the $q$-deformed case.

For an illustration, look at the simpler Painlev\`e III equation. The bilinear form of the equation, using a single $\tau$-function is given by the operator
\begin{equation}\label{PIII1}
    D^{I I I}=\frac{1}{2} D_{\log z}^{4}-z \frac{\mathrm{d}}{\mathrm{d} z} D_{\log z}^{2}+\frac{1}{2} D_{\log z}^{2}+2 z D_{\log z}^{0}:\ \ \ \ \ \ \ D^{I I I}\tau\cdot\tau=0
\end{equation}
Now if one introduces $\tau_1(z)$ related to $\tau(z)$ by a B{\"a}cklund transformation, one obtains the following form of the equation:
\begin{equation}\label{PIII2}
    \begin{aligned}
&D_{\log z}^{2}\tau\cdot \tau=-2 z^{1 / 2} \tau_{1}^{2} \\
&D_{\log z}^{2}\tau_{1}\cdot \tau_{1}=-2 z^{1 / 2} \tau^{2}
\end{aligned}
\end{equation}
The corresponding variable $w(z)$ which solves the Newtonian form of the Painlev\`e III equation is
\begin{equation}
    w(z) = \sqrt{z}\dfrac{\tau^2}{\tau_1^2}
\end{equation}
We discuss a similar representation of the Painlev\`e VI equation below, which uses 8 $\tau$-functions and is a continuous limit of (\ref{8qint})-(\ref{8qvir}).

\paragraph{Continuous limit of the q-Painlev\`e VI equation.}
Since the limit of the $q$-Virasoro conformal block is straightforward, we expect this limit to hold for all structures: the solution and the defining 8 equations. Surprisingly, we will see that the limiting procedure is not quite obvious and its relation with solution \eqref{PVIsigmasolution} is not that explicit.
\\\\
We have described above how the $q$-Painlev\`e VI equations are solved by the ratios of shifted $\tau$-functions, and one should expect that taking the continuous limit would provide a solution to the Painlev\`e VI equation.
The limit of the $q$-Painlev\`e equations (\ref{qPVI1})-(\ref{qPVI2}) themselves is rather tricky \cite{JS}. The procedure of taking the limit requires, first, defining the Hamiltonian variables $w(z),p(z)$ by the following limiting procedure:
\begin{equation}
    w(z) = w_1(z) ,\qquad \frac{\left(w_{1}-a_{1} z\right)\left(w_{1}-a_{2} z\right)}{\left(w_{1}-z\right)\left(w_{1}-1\right)} \frac{1}{q w_{2}}=1-\epsilon w_{1} p(z)
\end{equation}
and then expanding the parameters $a_1 = 1 +\epsilon \mathfrak{a}_1, b_1 =1+ \epsilon \mathfrak{b}_1, q \rightarrow 1+\epsilon $.  This reparametrization allows one to exclude $w_2(z)$ and hence to rewrite the $q$-Painlev\`e system in terms of $w(z),p(z)$. In the first order in $\epsilon$, one ends up with the Hamiltonian form of the Painlev\`e VI equation:
\begin{equation}
    \dfrac{dw}{dz} = \dfrac{\partial H_{VI}}{\partial p}, \qquad  \dfrac{dp}{dz} = -\dfrac{\partial H_{VI}}{\partial w}
\end{equation}
The Hamiltonian form is not easy to deal with, since the $p(z)$ function is quite complicated. Nevertheless, we expect the function $w(z)$ to solve the Newtonian form of the equation.

We can clearly do this with the solution given by \eqref{qPw}. Taking the limit in terms of the $\tau$-functions, we obtain the function
\begin{equation}\label{pvisol}
w(z) := \lim_{q \rightarrow 1} w_1(z)=z \dfrac{\hat{\tau}_1\hat{\tau}_2}{\hat{\tau}_3 \hat{\tau}_4}
\end{equation}
where $\hat{\tau}_i(z) = \lim\limits_{q \rightarrow 1} \tau_i(q,z)$, which should be a solution to the Painlev\`e VI equation in the Newtonian form.

The parameters of the obtained equation are of course related to the multiplicities $\alpha_i$. Interestingly enough, we get the Painlev\`e equation with parameters differing from those in \eqref{parameters}. For the solution obtained in the limit, we have:
\begin{equation}\label{parameters2}
\begin{split}
&\theta_1=\dfrac{1-N-2\alpha_2}{2}, \qquad \theta_t=\dfrac{-N-2\alpha_3}{2} \\
&\theta_0=\dfrac{-2\alpha_1-2\alpha_2-2\alpha_3-N}{2}
\qquad \theta_\infty+\frac{1}{2}=\dfrac{N+2\alpha_1}{2}
\end{split}
\end{equation}
These relations between the parameters of solutions differs from \eqref{parameters}. Therefore, in order to make a connection between the two representations, one should take the $\tau$-function form with the B{\"a}cklund transformed parameters:
\begin{equation}
   \tilde{\sigma}(z) =  z (z-1) \dfrac{d \log Z\left(-\alpha_1-\alpha_2-\alpha_3-N/2,-N/2-\alpha_3,\dfrac{1-N}{2}-\alpha_2 \right) }{d z}
\end{equation}
Then \cite{lisovyy2011dyson}:
\begin{equation}
    \begin{split}
\tilde{\sigma}(z)=& \frac{z^{2}(z-1)^{2}}{4 w(w-1)(w-z)}\left(w^{\prime}-\frac{w(w-1)}{z(z-1)}\right)^{2}-\frac{\theta_{0}^{2} z}{ w}+\frac{\theta_{1}^{2}(z-1)}{(w-1)}-\frac{\theta_{z}^{2} z(z-1)}{(w-z)} \\
&-\theta_{\infty}^{2}(w-1+z)
\end{split}
\end{equation}
with $w(z)$ given by formula \eqref{pvisol} and parameters given by \eqref{parameters2}. Equivalently, one could use formula \eqref{sigmatonewton} to invert the transformation.

\subsection{Continuous limit of the 8 bilinear equations}
One may also ask what is a continuum limit of the bilinear representation of the $q$-Painlev\`e system. An answer is given in \cite{BGramm}, which we review here more carefully. Just as the limit of (\ref{qPVI1})-(\ref{qPVI2}), it is not actually straightforward: one needs a special limiting procedure. Note that the parameter $q$ appears in the $\tau$-functions in two ways: as a parameter of the function and in difference shifts of the $z$-variable. We take a coarse continuous limit in the parameters of the $\tau$-functions, while keeping track of the $q$-dependence in the shifts:
\begin{equation}\label{taulimit}
     \tau(q, q^az) \rightarrow  \hat{\tau}(e^{a \epsilon}z)= \lim_{q \rightarrow 1} \tau(q, e^{a \epsilon}z)  \qquad \hbox{for}
     \qquad q \rightarrow \exp(\epsilon)
\end{equation}
To give an example, look at the $N=1$ case.  The natural $q \rightarrow 1$ limit would look like:
\begin{equation}
\begin{split}
Z^{(5d)}_1 &( \alpha_1,\alpha_2 ,\alpha_3,z)=Z^{(4d)}_1(\alpha_1,\alpha_2,\alpha_3,z)+ \\&+ \epsilon \left( \left( \alpha_1+\alpha_2+2\alpha_1 \alpha_2 \right) Z^{(4d)}_1(\alpha_1,\alpha_2,\alpha_3,z) - \left(\alpha_2+\alpha_3 + 1\right) 2\alpha_3 Z^{(4d)}_1(\alpha_1+1/2,\alpha_2,\alpha_3-1/2,z) \right)
\\
Z^{(5d)}_1&( \alpha_1,\alpha_2 ,\alpha_3,q z)=Z^{(4d)}_1(\alpha_1,\alpha_2,\alpha_3,z)+ \\&+ \epsilon \left( \left( \alpha_1+\alpha_2+2\alpha_1 \alpha_2 \right) Z^{(4d)}_1(\alpha_1,\alpha_2,\alpha_3,z) - \left(\alpha_2+\alpha_3 + 1\right) 2\alpha_3 Z^{(4d)}_1(\alpha_1+1/2,\alpha_2,\alpha_3-1/2,z)+ \right.
\\&
+ \left. z\dfrac{d}{dz} Z^{(4d)}_1(\alpha_1,\alpha_2,\alpha_3,z) \right)
\end{split}
\end{equation}
Instead, we make the following substitutions:
 \begin{equation}\label{Zlimit}
 \begin{split}
     Z^{(5d)}_1(\alpha_1,\alpha_2,\alpha_3,z) \  \longrightarrow& \ Z^{(4d)}_1(\alpha_1,\alpha_2,\alpha_3,z) \\ Z^{(5d)}_1(\alpha_1,\alpha_2,\alpha_3, qz)  \ \longrightarrow& \  Z^{(4d)}_1(\alpha_1,\alpha_2,\alpha_3,z)+ \epsilon \, z \dfrac{d}{dz} Z^{(4d)}_1(\alpha_1,\alpha_2,\alpha_3,z)
 \end{split}
 \end{equation}
The limit in \cite{BGramm} also involves expanding the coefficients $a_i,b_i$ only to a certain order in $\epsilon$, whereas it is unclear how to proceed with this method in our case, where the coefficients are explicit functions in $q$.

Thus, now we substitute (\ref{taulimit},\ref{Zlimit}) into the bilinear  equations (\ref{8qint},\ref{8qvir}) and expand them in $\epsilon$ up to the second order. However, only 8 of the resulting $8 \cdot 3$ equations correctly hold, and we just keep only them. Finally, we obtain a set of equations analogous to those in \cite{BGramm}:
\begingroup
\addtolength{\jot}{0.2em}
\begin{align}
 \epsilon^0: \qquad &z \htau_1 \htau_2 -z \htau_5 \htau_6- z\htau_3 \htau_4=0 \label{differatialhirota1}\\
&\tauh_3 \tauh_4+\tauh_7 \tauh_8 -z\tauh_1 \tauh_2=0\label{differatialhirota2} \\
\epsilon^1: \qquad &
(n-1+2\alpha_2) \tauh_3\tauh_4+\left(1-n+2\alpha_2+D_{\log z}\right)\tauh_7 \tauh_8=0\label{differatialhirota3} \\
& (n+2\alpha_3)\tauh_1\tauh_2+\left(-2\alpha_1-2\alpha_2-2\alpha_3+n-2-D_{\log z}\right)\tauh_5 \tauh_6=0
\label{differatialhirota4}\\
&(2\alpha_1+n)\tau_7\tau_8+\left(-2\alpha_2-2\alpha_3+(1-z)\left(2+2\alpha_1+2\alpha_2+2\alpha_3-n\right)+(1-z)D_{\log z}\right)\tauh_3 \tauh_4=0
\label{differatialhirota5}\\
&(-2\alpha_1-2\alpha_2-2\alpha_3-n)\tauh_5\tauh_6+\left(2\alpha_3+3n-2+z(2+2\alpha_2-3n)-z(1-z)D_{\log z}\right)\tauh_1\tauh_2=0
\label{differatialhirota6}\\
\epsilon^2: \qquad &
\tauh_1 \tauh_2 \psi_{12}\left( \vec{\alpha},N,z\right)+\tauh_7 \tauh_8 \psi_{78}\left( \vec{\alpha},N,z\right)
   =\nonumber\\&
=-(1-z)^2 D^2_{\log z} \tauh_3\tauh_4-(1-z)zD^2_{\log z}\tauh_5\tauh_6+(1-z)D^2_{\log z}\tauh_7 \tauh_8 \label{differatialhirota7} \\\nonumber
&\\ \nonumber&
\tauh_3 \tauh_4 \psi_{34}\left( \vec{\alpha},N,z\right)+\tauh_5 \tauh_6 \psi_{56}\left( \vec{\alpha},N,z\right)
=
\nonumber\\&=-(1-z)^2 D^2_{\log z} \tauh_1\tauh_2-(1-z)zD^2_{\log z}\tauh_5\tauh_6+(1-z)D^2_{\log z}\tauh_7 \tauh_8 \label{differatialhirota8}
\end{align}
\endgroup
Here$\psi_{ij}\left( \vec{\alpha},N,z\right)$ are quadratic polynomials in all the variables. They are rather complicated and are ambiguously determined because of relations (\ref{differatialhirota1},\ref{differatialhirota2}).
These bilinear equations are another bilinear form of the Painlev\`e VI equation. Instead of a single equation of order 4, one has a system of lower order equations for 8 $\tau$-functions.

These bilinear equations are equivalent to the Newtonian form of the Painlev\`e VI equation for the function
\begin{equation}\label{ratioPVI}
    w(z)=z \dfrac{\hat{\tau}_1\hat{\tau}_2}{\hat{\tau}_3 \hat{\tau}_4}
\end{equation}
Note that, due to equations \eqref{differatialhirota1},\eqref{differatialhirota2}, various representation of $w(z)$ are allowed, for example: $w(z)=1+\dfrac{\hat{\tau}_7\hat{\tau}_8}{\hat{\tau}_3 \hat{\tau}_4}$.

Now looking at these equations from the point of view of the eigenvalue integral, one can again distinguish between those equations that depend on the measure $\xi(dx)$ and those which do not. The first two equations \eqref{differatialhirota1}-\eqref{differatialhirota2} are certainly nothing but the Hirota equations \eqref{hirotamiwa}.

As in the $q$-deformed case, we expect that in total we have four equations that we attribute to integrability. The missing two are constructed as linear combinations, and we obtain:
\begin{equation}\label{IntegrHir}
\begin{split}
        &z \htau_1 \htau_2 -z \htau_5 \htau_6- z\htau_3 \htau_4=0 \\
&\tauh_3 \tauh_4+\tauh_7 \tauh_8 -z\tauh_1 \tauh_2=0\\
\eqref{differatialhirota3} + \eqref{differatialhirota4} -\eqref{differatialhirota6}&=(2n-1)\tauh_1\tauh_2+(1-z)D_{\log z}\tauh_1\tauh_2-z D_{\log z} \tauh_5 \tauh_6-D_{\log z} \tauh_7 \tauh_8=0 \\
   \eqref{differatialhirota3} + z\eqref{differatialhirota4} -\eqref{differatialhirota5}&=-z \tauh_1 \tauh_2 +(1-z)D_{\log z}\tauh_3\tauh_4-z D_{\log z}\tauh_5 \tauh_6+D_{\log z}\tauh_7 \tauh_8=0
    \end{split}
\end{equation}
The other four equations are consequence of Virasoro-like constraints. It is not clear how to generally identify the full derivative insertions that correspond to these equations. One can do it in particular cases like, for example, for $N=2$. However, these explicit expressions are not too illuminating, and we will not provide them here.

All in all, we see that the structure behind the relation between the Painlev\`e equation and the integrability+string equations survives the continuous limit. While now it is made complicated by distinct forms of the PVI equations and by the peculiarities of the continuous limit. In fact, one gives a slightly different interpretation of equations \eqref{8qint}-\eqref{8qvir} as a consistency condition between representation \eqref{ratioPVI} and \eqref{PVIsigmasolution}. Formula \eqref{sigmatonewton} is then equivalent (up to B{\"a}cklund transformations) to the system of 8 bilinear equations.

Finally, notice that the naive substitution of $q=1$ into the $q$-Painlev\`e equations leads to:
\begin{equation}
\begin{split}
        w_1^2&= \left( \dfrac{w_2-z}{w_2-1} \right)^2
        \\
        w_2^2&= \left( \dfrac{w_1-z}{w_1-1} \right)^2
\end{split}
\end{equation}
which are consistent with the equations (\ref{differatialhirota1},\ref{differatialhirota2}), for example, the second one is rewritten as:
\begin{equation}
    w_2 = \dfrac{w_1-z}{w_1-1} \Longrightarrow z\dfrac{\tauh_5 \tauh_6}{\tauh_7 \tauh_8} = \dfrac{\tauh_1 \tauh_2-z \tauh_3\tauh_4}{\tauh_1 \tauh_2 -\tauh_3 \tauh_4}
 \end{equation}

\section{Pure gauge theory limit}
\subsection{The Painlev\`e III equation}

It appears that the described properties persist in the pure gauge limit of the AGT correspondence in $4d$. In this limit, the Nekrasov partition functions and the conformal blocks are described by the  Brez\'in-Gross-Witten (BGW) matrix model. On the gauge theory side, one has an $\mathcal{N}=2$ pure gauge theory, which is no longer conformal. The masses of hypermultiplets are set to infinity, and a scale parameter $\Lambda$ emerging due to renormalizations is introduced. In terms of parameters of the conformal block (which is called irregular conformal block in this limit \cite{irregCB}), this limit is given by
\begin{equation}
\begin{split}
\alpha_i& \rightarrow 0 , \, \ \ \  z \rightarrow \infty \\
(\alpha_1^2-\alpha_2^2)&(\alpha_3^2-\alpha_4^2)z \ \ \  \text{   fixed}
\end{split}
\end{equation}
The pure gauge limit corresponds to the reduction from the Painlev\`e VI to Painlev\`e III$_3^'$ equation. It is clear that after the limit is taken there are no parameters left in the equation. Just as the Painlev\`e VI equation, it has several forms. In particular, the Newtonian form is given by
\begin{equation}
    \frac{d^{2} w}{d z^{2}}=\frac{1}{w}\left(\frac{d w}{d z}\right)^{2}-\frac{1}{z} \frac{d w}{d z}+\frac{2 w^{2}}{z^{2}}-\frac{2}{z}
\end{equation}
while the $\sigma$-form is
\begin{equation}
    \left(z \sigma^{\prime \prime}\right)^{2}=4\left(\sigma^{\prime}\right)^{2}\left(\sigma-z \sigma^{\prime}\right)-4 \sigma^{\prime}
\end{equation}
We have already presented the bilinear forms of the equation (\ref{PIII1}) and (\ref{PIII2}), the second one uses two $\tau$-functions related by the B{\"a}cklund transformation
\begin{equation}\label{P3hirota}
    D^2_{\log z} \tau^2= -2z^{1/2}\tau_1^2
\end{equation}
Transformations between different representations are rather simple in this case:
\begin{equation}\label{PIIIrel}
    w^{\text{III}}= -\dfrac{1}{( \sigma^{\text{III}})'} , \qquad w =z^{1/2}\dfrac{\tau^2}{\tau_1^2} , \qquad \sigma^{\text{III}}= z \dfrac{d \log \tau}{z}
\end{equation}
In fact, it is rather easy to see consistency between the Painlev\`e equations and the transformation formulas \eqref{PIIIrel}. For instance, substituting expressions for $\sigma$ and $w$ into the first identity in \eqref{PIIIrel}, one immediately obtains the bilinear equation \eqref{P3hirota}.

Various objects of the Painlev\`e VI theory have specific scaling properties in the pure gauge limit. The scaling of the Painlev\`e VI variables to the Painlev\`e III ones is given by:
\begin{itemize}
    \item Newtonian variable:
    \begin{equation}
            w^{\text{III}}(z) = \lim_{\alpha_i \rightarrow 0 }   \left[ \dfrac{z (\alpha_1^2-\alpha_2^2)}{(\alpha_3+\alpha_3)} \cdot  w^{\text{VI}}\left(  \dfrac{z}{(\alpha_1^2-\alpha^2)(\alpha_3^4-\alpha_4^2)}\right)  \right]
            \end{equation}
    \item $\tau$-function:
    \begin{equation}
        \tau^{\text{III}}= \lim_{\alpha_i \rightarrow 0 }   \left[ \left(\dfrac{z}{(\alpha_3+\alpha_4)}\right)^{\alpha_1^2+\alpha_2^2} \cdot  \tau^{\text{VI}}\left(  \dfrac{z}{(\alpha_1^2-\alpha^2)(\alpha_3^4-\alpha_4^2)}\right)  \right]
    \end{equation}

\end{itemize}

The scaling properties of the eight $\tau$-functions are not evident because of the shift structure. We propose an answer from analyzing the matrix model representation in the next subsection.

\subsection{PGL conformal block as BGW matrix model}
In the pure gauge limit, the conformal block is given by BGW integrals as follows. Consider the BGW partition function \cite{GMU} given by the integral over unitary $N\times N$ matrix $U$ and depending on the external matrix $\Psi$:
\begin{equation}
    Z_{B G W}(N \mid \Psi)=\frac{1}{{\rm Vol}_N} \int[d U] e^{\left(\operatorname{Tr} U^{\dagger}+\operatorname{Tr} \Psi U\right)}
\end{equation}
and define the partition function, which is a matrix model realization of the irregular conformal block corresponding to the Nekrasov functions in pure gauge theory (with dimensional parameter $z=\Lambda^4$, since the gauge theory is no longer conformal) \cite{MMShBGW,MMP}:
\begin{equation}\label{Z1}
\begin{split}
Z_{*}^{(1)}(N,z)=&\int[d U] \int[d V] Z_{B G W}\left(m_{+} \mid U\right) Z_{B G W}\left(m_{-} \mid V\right) \operatorname{det}\left(1-z U^{\dagger} \otimes V^{\dagger}\right)^{2}= \\
=& \sum_{R,Q} (-\Lambda^4)^{|R|} \delta_{|R|,|Q|}\cdot K_{RQ}
\cdot \frac{d_R^2d_Q^2}{D_{_R}(m_+) D_{_Q}(m_-)}
\end{split}
\end{equation}
where
\be
K^{RQ} = \sum_{X,Y} {\cal N}^{R}_{X,Y}{\cal N}^{Q}_{X^{tr},Y^{tr}}
\ee
and ${\cal N}^R_{PQ}$ are the Littlewood-Richardson coefficients, $D_R(m)$ is the dimension of representation $R$ of group $SL(m)$, $d_R$ is the dimension of representation $R$ of symmetric group $S_{|R|}$ \cite{Book}.
The partition function \eqref{Z1} depends on sizes of matrices $U$ and $V$, which are chosen to be $m_+=N$ and (analytically continued) $m_-=-N$.

As in the previous cases, the integrability properties and the relation to the Painlev\`e equation is revealed after the Fourier transform:
\begin{equation}\label{zpgldef}
    Z_{PGL}\left(a,z\right)=\sum_{k \in \mathbb{Z}} \mathfrak{Z}(a+k,z) e^{i k \eta} , \quad \mathfrak{Z}(a,z)=\frac{z^{a^{2}} Z_{*}^{(1)}(2 a,z)}{\mathfrak{G}(1+2 a) \mathfrak{G}(1-2 a)}
\end{equation}
where $\mathfrak{G}(x)$ is the Barnes $G$-function \cite{Barnes}, and we have chosen this normalization so that $Z_{PGL}\left(a,z\right)$ would be equal to the $\tau$-functions of the Painlev\`e III equation with respect to the variable $z$, while the B{\"a}cklund transformed $\tau$-function in \eqref{P3hirota} is given by a shift in the $a$ parameter:
\begin{equation}
    \tau^{\mathrm{III}}= Z_{PGL}\left(a,z\right) \, , \qquad \tau^{III}_{1} = Z_{PGL}\left(a+\dfrac{1}{2},z\right)
\end{equation}
One can understand this shift as a remnant from the Painlev\`e VI $\tau$-functions as follows. The parameter $a$ is the internal dimension in the conformal block, i.e. it is a remnant of $\alpha$ in the pure gauge limit, for which we remind the relation: $N_1=\alpha+\alpha_1+\alpha_2 \, \  N_2=-\alpha+\alpha_3+\alpha_4$. The shifts of $\alpha_i$ parameters in the 8 $\tauh_i$ functions are suited so that, for $i=1,2,7,8$, they can be thought of as effectively shifting $\alpha \rightarrow \alpha+\frac{1}{2}$.

As these formulas suggest, the scaling limit from the 8 Painlev\`e VI $\tau$ functions to the Painlev\`e III $\tau$-functions sends:
\begin{equation}
    \tauh_{i}  \longrightarrow  \left\{\begin{split}
        &\tau^{(III)} \, , \ i=3,4,5,6
        \\
        &\tau_1^{(III)}\, , \ i=1,2,7,8
    \end{split}
    \right.
\end{equation}
Just as in the previous case, there are various forms of the Painlev\`e III equation in terms of the PGL conformal blocks:
\begin{equation}
    \sigma^{\text{III}}(z)= z \dfrac{d }{dz} Z_{PGL}\left(a,z\right), \qquad w^{\text{III}}(z)=z^{1/2} \dfrac{Z_{PGL}\left(a,z\right)^2}{Z_{PGL}\left(a+\dfrac{1}{2},z\right)^2}
\end{equation}
\paragraph{Painlev\`e III as a reduction of the Toda equation.}
The BGW description of the PGL conformal block implies that is satisfies the Toda system:
\begin{equation}
  D^2_{\log z}Z_{PGL}(a,z)\cdot Z_{PGL}(a,z)=-2 z^{1/2}Z_{PGL}(a-1/2,z)Z_{PGL}(a+1/2,z)
\end{equation}
with $n=2a$ playing the role of the discrete Toda time. However, due to the structure of \eqref{zpgldef}, it obeys the periodicity condition
\begin{equation}
    Z_{PGL}(a+1,z)=Z_{PGL}(a,z)
\end{equation}
This condition plays the role of a reduction constraint that reduces the Toda equations to the Painlev\`e III equation, which acquires the bilinear form \eqref{P3hirota}:
\begin{equation}
    D_{\log z}^2 Z_{PGL}(a,z)= -2 z^{1/2} Z_{PGL}(a+1/2,z)^2
\end{equation}

\section{Discussion}

To conclude, we provided a detailed description of the relation between conformal blocks realized in terms of
matrix models and the Painlev\`e $\tau$-functions (see \cite{BGT} for a different relation of $q$-Painlev\'e with matrix models, see also \cite{GG}).
We gave a uniform presentation of both the $q$-deformed case ($5d$ theories)
and the non-deformed one ($4d$ theories).
We explained how the relevant limit $q\longrightarrow 1$ is taken and
how the set of $8$ Painlev\`e $\tau$-functions gets split into Hirota bilinear identities for the ordinary KP $\tau$-functions
and Virasoro-like constraints, the two basic ingredients of the standard theory of (eigenvalue) matrix models \cite{UFN3}.

What remains to be done besides a further work on clarification of above ideas
is to understand a relation of Painlev\`e theory to the modern view on eigenvalue matrix models
based on the phenomenon of superintegrability (see \cite{MMsi} and references therein).
At least for fixed integration contours, one could use the integrable structure and the string equation to convert all equation into pure algebraic/combinatoric ones and solve them for obtaining an explicit answer. An even more straightforward way of solving the Virasoro constraints using the so-called $W$-representation is now under development \cite{wrep,wrep2,MSh,Alex,Max,MMMs}. Superintegrability is effectively applied \cite{MMNek} to logarithmic matrix models which we considered in this paper. We hope that these problems will attract attention of the community,
and plan to return to them elsewhere.

\section*{Acknowledgments}

This work was supported by the Russian Science Foundation (Grant No.20-12-00195).


\end{document}